\documentclass{PoS}
\usepackage{amssymb,amsmath,array}

\title{Angular momentum sum rule in nuclei}

\ShortTitle{Angular momentum sum rule in nuclei}
 
\author{Swadhin Taneja \\  Stony Brook University - Department of Physics and Astronomy \\
Stony Brook, NY 11776 - USA \\ E-mail: \email{taneja@skipper.physics.sunysb.edu}}

\author{Simonetta Liuti \\ University of Virginia - Physics Department \\
382, McCormick Rd., Charlottesville, Virginia 22904 - USA \\ 
E-mail: \email{sl4y@virginia.edu}}
\abstract{In this work we derive a sum rule for the angular momentum of a spin 1 hadronic system.}

\FullConference{XVIII International Workshop on Deep-Inelastic Scattering and Related Subjects\\
                April 19 -23, 2010\\
                Convitto della Calza, Firenze, Italy}

\begin{document}
\section{Introduction}
One of the outstanding questions in QCD is  the proton spin puzzle. A number of experiments performed
since the '80s, including the most recent HERMES, Jefferson Lab and Compass measurements,  
have confirmed that only about $30 \%$ of the proton spin is accounted by quarks, and that the quark contribution
is dominated by the valence component. Current efforts, both in theory and experiment, are therefore directed towards
determining the contributions of the  Orbital Angular Momentum (OAM) of the quarks, as well as of the spin and 
OAM of the gluons. 
There exist two essential ways of defining sum rules for the proton's total angular momentum based on different
gauge treatment and quantization choices. These were proposed by Jaffe and Manohar (JM) \cite{Jaffe:1989jz}  and by X. Ji (Ji) \cite{Ji:1996ek}, respectively.
\footnote{Alternative procedures where explicit gauge invariant operators for spin and orbital angular momentum of quarks and gluons are claimed to be obtained were given in \cite{Bakker:2004ib,Chen:2008ag}. However their discussion is beyond the scope
of this contribution.}
The two approaches imply different 
experimental access to the angular momentum components. On one side, with JM's decomposition one can measure the quarks spin from {\it e.g.} Deep Inelastic Scattering (DIS), and the gluon spin from both DIS and proton-proton collisions. On the other, Ji's decomposition, where all terms are manifestly gauge invariant, allows one to determine the quarks OAM from Deeply Virtual Compton Scattering (DVCS) type experiments. By considering a QCD factorization based description of such processes one in fact obtains an additional sum rule,
\begin{equation}
\label{Ji}
J_q = \frac{1}{2} \int d x \, x \left[ H_q(x,0,0) + E_q(x,0,0) \right],
\end{equation}    
where we have introduced the Generalized Parton Distributions (GPDs), $H_q$ and $E_q$.

Experimental extractions of the quark total angular momentum through Ji's sum rule, Eq.(\ref{Ji}), have been given by several groups \cite{:2008jga,:2007vj} 
in the form of ``model dependent" constraints on $J_u$ vs.  $J_d$. Model dependence is  introduced because  various theoretical models using $J_u$ and $J_d$ as parameters are used to fit the exclusive data on both DVCS \cite{:2007vj} and on the transverse target spin asymmetry \cite{:2008jga}. It was, however, claimed in \cite{:2008jga} that the total propagated experimental uncertainty "dominates the effects of variations within the GPD models".     

Motivated by the challenge of the spin puzzle on one side, and by the experimental progress on the other, 
we started a program to investigate OAM in nuclear targets. 
In this contribution we present initial results on the total angular momentum in a spin one nucleus, 
namely deuterium. It should be in fact also mentioned that  analyses of 
deeply virtual exclusive processes on nuclei are presently  ongoing at Jefferson Lab \cite{4He}, and that
nuclear targets will be considered in the Lab's 12 GeV upgrade. 

The new sum rule reads
\begin{equation}
\label{Ji_deuteron}
J_{q}=\frac{1}{2} \int dx \, x \, H_{2}^{q}(x,0,0),   
\end{equation}
where $H_{2}^{q}(x,\xi,t)$ is the deuteron GPD corresponding to the form factor $G_2(t) \equiv G_M(t)$,
{\it i.e.} the magnetic form factor. Similarly to $E$, the GPD $H_2$, does not have a forward limit \cite{Berger:2001zb}.
In what follows we outline the fundamental steps of the derivation. Further phenomenological developments including 
nuclear effects, as well as procedures to extract the deuteron's total angular momentum from experiment will be presented in an upcoming publication.    

\section{QCD decomposition of proton spin}
\label{sec2}
We start be describing the JM \cite{Jaffe:1989jz}  and Ji \cite{Ji:1996ek}
angular momentum decompositions,
\begin{eqnarray}
\frac{1}{2} = \sum_q J_q + J_g,
\end{eqnarray}
where in Ref.\cite{Jaffe:1989jz} 
\begin{eqnarray}
\label{JMq}
J_q^{JM}  & = &  \int d^3 x \: \psi^\dagger \left[ \frac{1}{2} \Sigma + {\bf x} \times (-i {\bf \nabla}) \right] \psi =  \frac{1}{2} \Delta q +  {\cal L}_q \\
\label{JMg}
J_g^{JM}  & = & \int d^3 x \left[ ({\bf E} \times {\bf A}  - E_i ({\bf x} \times {\bf A}) A_i \right] =  \Delta G + {\cal L}_g,
\end{eqnarray}
while in Ref.\cite{Ji:1996ek},
\begin{eqnarray}
\label{Jiq}
J_q^{Ji}  =  \int d^3 x \psi^\dagger \left[ \frac{1}{2} \Sigma + {\bf x} \times (-i {\bf \partial} - g {\bf A}) \right] \psi =  \frac{1}{2} \Delta q + \sum_q  L_q;  \; \; 
J_g^{Ji}   =  \int d^3 x \left[ {\bf x} \times ({\bf E} \times {\bf B}) \right] 
\end{eqnarray}
Eqs.(\ref{JMq},\ref{JMg},\ref{Jiq}) 
can be derived by considering the angular momentum tensor in QCD  \cite{Ji:1996ek}, 
\begin{equation}
     J^i = \frac{1}{2}\epsilon^{ijk} 
      \int d^3x M^{0jk} \ ,
\label{e:angmom}       
\end{equation}
where $M^{0ij}$ is the angular momentum density given in terms of the symmetric, gauge-invariant, and conserved energy momentum tensor as $
      M^{\alpha\mu\nu} = T^{\alpha\nu} x^\mu - T^{\alpha\mu} x^\nu$. 
$T^{\mu\nu}$ has separate gauge invariant contributions from quarks and gluons,
\begin{equation}
  T^{\mu\nu} =   T^{\mu\nu}_q + T^{\mu\nu}_g 
             =   \frac{1}{2}[\bar \psi \gamma^{(\mu} i
      \overrightarrow{ D^{\nu)}} \psi + 
       \bar \psi \gamma^{(\mu} i
           \overleftarrow {D^{\nu)} } \psi] +
           \frac{1}{4}g^{\mu\nu}
       F^2 - F^{\mu\alpha}F^\nu_{~\alpha} \ 
\label{e:angmomt}       
\end{equation}


%

The connection of GPD's to the angular momentum becomes apparent by first writing down the form factors of the quarks and gluon part of the QCD energy-momentum tensor, $T^{\mu\nu}_{q,g}$ , 
\begin{equation}     
      \langle p'| T_{q,g}^{\mu\nu} |p\rangle
       = \bar U(p') \Big[A_{q,g}(t)
       \gamma^{(\mu} P^{\nu)} +
   B_{q,g}(t) P^{(\mu} i\sigma^{\nu)\alpha}\Delta_\alpha/2M
  +  C_{q,g}(t)\Delta^{(\mu} \Delta^{\nu)}/M \Big] U(p)\ .
\label{e:angmompro}
\end{equation}   
Using Eqs.(\ref{e:angmom}) and (\ref{e:angmompro}) one can derive the quark and gluon total angular momentum contribution in terms of form factors of $T^{\mu\nu}_{q,g}$ \cite{Ji:1996ek},
\begin{eqnarray}
      J_{q, g} = {1\over 2} \left[A_{q,g}(0) + B_{q,g}(0)\right] \ . 
\label{e:angmomqg}
\end{eqnarray} 
Cleary, measuring directly the gravitational form factors, $A_{q,g}$ and $B_{q,g}$, would be a hopeless task.
However one can match these form factors with the amplitude for (unpolarized) DVCS  
\begin{eqnarray}
& & \int {d\kappa\over 2\pi} e^{ix \kappa P.n}
    \left\langle p'\left|\overline \psi_q \left(-{\kappa}n\right)
      \not \! n  
    \psi_q\left({\kappa}n\right) \right| p\right\rangle =   \nonumber \\
 & &  \frac{1}{P.n}\left[H_q(x, \xi, t) \overline U(p')\not\! n U(p)  
     +  E_q(x, \xi, t) \overline U(p') {i\sigma^{\mu\nu}
  n_\mu \Delta_\nu \over 2M} U(p)\right] \ ,
\label{e:GPDs}
\end{eqnarray}  
where $H_q$ and $E_q$ are the quark GPDs. 
At leading twist (i.~e.~twist 2) and for a spin-1/2 target such as the proton, there are four GPDs needed for chiral-even processes, {\it i.e.} processes that do not flip the chirality of the involved quark. Another four GPDs that are chiral-odd appear in processes related to transversity. 
$H_q$ and $E_q$, are functions of the longitudinal momentum, $x$, a skewness variable, $\xi = -\frac{\Delta.n}{P.n}$ and invariant momentum transfer, $t = {\Delta}^2$, where $P=p+p'$ and $\Delta=p'-p$.

It follows that  by expanding the matrix element on the left hand side of equation (\ref{e:GPDs}) and taking the second moment with respect to $x$ one can find the following relation between $A_{q,g}$,  $B_{q,g}$ and $H_{q,g}$, $E_{q,g}$, 
\begin{equation}
     \int dx x [H_{q,g}(x, \xi, t) +
       E_{q,g}(x, \xi, t) ]
     = A_{q,g}(t) + B_{q,g}(t) 
\label{e:GPDstenFF}     
\end{equation}

In the limit $t \rightarrow 0$ then one finds a relation between $H_{q,g}$, $E_{q,g}$ and $J_{q, g}$, as defined in equation (\ref{Ji}),
\begin{equation}
     J_{q, g}=\frac{1}{2}\int dx x [H_{q,g}(x, \xi, 0) +
       E_{q,g}(x, \xi, 0) ] 
\label{e:GPDsumpro}     
\end{equation}  

\section{Deuteron generalized parton distributions and spin sum rule}
\label{sec3}
As a natural extension of studies of the substructure of deuteron using parton distributions \cite{Hoodbhoy:1988am}, Berger {\it et al.} studied generalized parton distributions in the deuteron \cite{Berger:2001zb}. Deuteron being a spin 1 system, there are in all five unpolarized GPDs,
\begin{eqnarray}
&&  \int \frac{d \kappa}{2 \pi}\,
  e^{i x \kappa P.n}
  \langle p', \lambda' |\,
  \bar{\psi}(-\kappa  n)\, \gamma.n\, \psi(\kappa n)
  \,| p, \lambda \rangle 
    =  - (\epsilon'^* .\epsilon) H_1 
+ \frac{(\epsilon .n) (\epsilon'^* .P)+ (\epsilon'^* .n) (\epsilon .P)}{P.n} H_2  \nonumber \\ 
&& - \frac{(\epsilon .P)(\epsilon'^* .P)}{2 M^2} H_3 
+ \frac{(\epsilon .n) (\epsilon'^* .P)- (\epsilon'^* .n) (\epsilon .P)}{P.n} H_4 
+ \Big\{4 M^2\, \frac{(\epsilon .n)(\epsilon'^* .n)}{(P.n)^2}+\frac{1}{3} (\epsilon'^* .\epsilon) \Big\}H_5 
\label{e:GPDsDeu}
\end{eqnarray}
where $n$ is a light-like vector and $\epsilon, \epsilon'$ are the polarization vector of deuteron in initial 
and final helicity state respectively.
As in the case of the nucleon we start out with writing down the energy momentum tensor, for our spin 1 system,
\begin{eqnarray}
\langle p' |T^{ \mu \nu} |p \rangle=
& - &  \frac{1}{2} \left[P^{\mu}P^{\nu} 
\right](\epsilon'^*  \epsilon)G_{1,2}(t) - \frac{1}{4} \left[P^{\mu}
P^{\nu} \right]
\frac{(\epsilon P)(\epsilon'^* P)}{M^2} G_{2,2}(t) \\ \nonumber
& - &  \frac{1}{2} \left[\Delta^{\mu} \Delta^{\nu} - g^{\mu \nu}
{\Delta^2}\right](\epsilon'^*  \epsilon)G_{3,2}(t) - \frac{1}{4}
\left[\Delta^{\mu}
\Delta^{\nu} - g^{\mu \nu}{\Delta^2}\right]
\frac{(\epsilon P)(\epsilon'^* P)}{M^2} G_{4,2}(t) \\ \nonumber
& + & \frac{1}{4} \left[ \left(\epsilon'^{* \mu} (\epsilon P)
 + \epsilon^{\mu} (\epsilon'^* P) \right) P^{\nu} +
\mu \leftrightarrow \nu \right]G_{5,2}(t) \\ \nonumber
& + & \left[ \left(\epsilon'^{* \mu} (\epsilon P)
 -  \epsilon^{\mu} (\epsilon'^* P) \right) \Delta^{\nu} +
\mu \leftrightarrow \nu  \right. \nonumber \\  & + & 
\left. 2 g^{\mu \nu} (\epsilon P)
(\epsilon'^* P) - \left(\epsilon'^{* \mu} \epsilon^{\nu}    + 
 \epsilon'^{* \nu} \epsilon^{\mu} \right) \Delta^2 \right] G_{6,2}(t)
\label{e:angmomdeu}
\end{eqnarray}

The above symmetric, gauge invariant tensor has six independent form factors, $G_{i,2}(t)$ where $i$ = 1-6. From equations (\ref{e:angmom}), and (\ref{e:angmomdeu}) then we get the quark and gluon total angular momentum contributions to the deuteron in terms of form factors of $T^{\mu\nu}_{q,g}$,
\begin{equation}
J_{q, g}=\frac{1}{2} G_{5,2}(0)   
\end{equation}
As in the nucleon case we expand the matrix element on the left hand side of equation (\ref{e:GPDsDeu}) and take the second moment with respect to $x$ and find the following relations between the GPDs $H$ and form factors $G$, 
\begin{eqnarray}
\int dx x H_{1}(x,\xi,t) - \frac{1}{3} \int dx x H_{5}(x,\xi,t) & = & G_{1,2}(t) + {\xi}^2 G_{3,2}(t) \\ \nonumber
\int dx x H_{2}(x,\xi,t) & = & G_{5,2}(t)  \\ \nonumber
\int dx x H_{3}(x,\xi,t) & = & G_{2,2}(t) + {\xi}^2 G_{4,2}(t) \\ \nonumber
\frac{1}{4 \xi} \int dx x H_{4}(x,\xi,t) & = & \frac {M^2}{t} \int dx x H_{5}(x,\xi,t)
=  G_{6,2}(t)
\label{e:GPDFFdeu}
\end{eqnarray}
In the limit $t \rightarrow 0$ then one finds a sum rule relation between the deuteron GPD $H_2$, and the angular momentum $J_{q, g}$, as defined in equation (\ref{Ji_deuteron}),
\begin{equation}
J_{q, g}=\frac{1}{2} \int dx \, x \, H_{2}^{q,g}(x,\xi,0).  
\end{equation}

\section{Conclusions}
In this work we derived a sum rule relation for the angular momentum contributions of quarks and gluons to a spin 1 system, {\it e.g.} the deuteron. We find that the second moment of the GPD $H_{2}^{q,g}$ is responsible for the angular momentum of the spin 1 system. It must be noted that $H_{2}$ is the same GPD whose first moment gives the magnetic moment of the deuteron ($G_{M}(t)$=$\int dx H_{2}^{q,g}(x,\xi,t)$).      


 

\section{Bibliography}
\begin{footnotesize}

\end{footnotesize}


\begin{thebibliography}{99}




\bibitem{url} Slides: \\ 
\verb$http://indico.cern.ch/contributionDisplay.py?sessionId=9&contribId=215&confId=86184$
\bibitem{Jaffe:1989jz}
  R.~L.~Jaffe and A.~Manohar,
  Nucl.\ Phys.\  B {\bf 337}, 509 (1990).
  
\bibitem{Ji:1996ek}
  X.~D.~Ji,
  Phys.\ Rev.\ Lett.\  {\bf 78}, 610 (1997)
  [arXiv:hep-ph/9603249].
  
 \bibitem{Bakker:2004ib}
  B.~L.~G.~Bakker, E.~Leader and T.~L.~Trueman,
  Phys.\ Rev.\  D {\bf 70}, 114001 (2004)

\bibitem{Chen:2008ag}
  X.~S.~Chen, X.~F.~Lu, W.~M.~Sun, F.~Wang and T.~Goldman,
  Phys.\ Rev.\ Lett.\  {\bf 100}, 232002 (2008)
  [arXiv:0806.3166 [hep-ph]].


\bibitem{:2008jga}
  A.~Airapetian {\it et al.}  [HERMES Collaboration],
  JHEP {\bf 0806}, 066 (2008)

\bibitem{:2007vj}
  M.~Mazouz {\it et al.}  [Jefferson Lab Hall A Collaboration],
  Phys.\ Rev.\ Lett.\  {\bf 99}, 242501 (2007)

\bibitem{4He} A. El Alaoui, http://meetings.aps.org/link/BAPS.2010.APR.S7.3. 


\bibitem{Berger:2001zb}
  E.~R.~Berger, F.~Cano, M.~Diehl and B.~Pire,
  Phys.\ Rev.\ Lett.\  {\bf 87}, 142302 (2001)
  
\bibitem{Hoodbhoy:1988am}
  P.~Hoodbhoy, R.~L.~Jaffe and A.~Manohar,
  Nucl.\ Phys.\  B {\bf 312}, 571 (1989).
  
  
    
 
\end{thebibliography}
\end{document}